\documentstyle[aps,12pt]{revtex}
\newcommand{\bb}{\begin{equation}}
\newcommand{\ee}{\end{equation}}
\newcommand{\ba}{\begin{array}}
\newcommand{\ea}{\end{array}}

\begin {document}
\baselineskip 2.2pc

\title{Bound states of neutral particles in external electric fields
\thanks{published in Phys. Rev. A {\bf 61} (2000) 022101}}
\author{Qiong-gui Lin\thanks{E-mail addresses: 
        qg\_lin@163.net, stdp@zsu.edu.cn}}
\address{China Center of Advanced Science and Technology (World
	Laboratory),\\
        P.O.Box 8730, Beijing 100080, People's Republic of China\\
        and\\
        Department of Physics, Zhongshan University, Guangzhou
        510275,\\
        People's  Republic of China \thanks{Mailing address}}

\maketitle
\vfill

\begin{abstract}
{\normalsize Neutral fermions of spin $\frac 12$ with magnetic
moment can interact with electromagnetic fields through nonminimal
coupling. The Dirac--Pauli equation for  such a fermion coupled to a
spherically symmetric or central electric field can  be reduced
to two simultaneous ordinary differential equations by separation
of variables in spherical coordinates. For a wide variety of central
electric fields, bound-state solutions of critical energy values
can be found analytically. The degeneracy of these energy levels
turns out to be numerably infinite. This reveals the possibility of
condensing infinitely many fermions into a single energy level.
For radially constant and radially linear electric fields, the system
of ordinary differential equations can be completely solved, and all
bound-state solutions are obtained in closed forms. The radially
constant field supports scattering solutions as well.
For radially linear fields, more
energy levels (in addition to the critical one)
are infinitely degenerate. The
simultaneous presence of central
magnetic and electric fields is discussed.}
\end{abstract}
\vfill
\leftline {PACS number(s): 03.65.Pm}
\newpage

\baselineskip 15pt

\section{Introduction}               

In relativistic quantum theory, a charged fermion of spin
$\frac 12$ moving in a background electromagnetic field is described
by the Dirac equation with minimal coupling to the vector potential.
In 1941, Pauli extended this equation to include an additional
nonminimal coupling term which takes into account the interaction
caused by  the anomalous magnetic moment of the charged particle [1].
This extended equation is usually called the Dirac--Pauli equation.
Many works have been devoted to the investigation of exact solutions
of this equation in various electromagnetic fields, say, a constant
uniform magnetic field, an electromagnetic plane wave, and more
complicated ones [2]. A constant central (spherically symmetric)
electric field was also considered by some authors. In this case the
Dirac--Pauli equation is separable in spherical coordinates, however,
exact solutions of the radial equation have not been  found in
closed forms [2]. In these studies, the nonminimal coupling is
conceptually taken as some correction to the minimal one (though the
correction is considerable for protons), and the simultaneous presence
of both couplings causes some mathematical difficulty.

In this paper we consider neutral fermions of spin $\frac 12$ with
magnetic moment. Without electric charges, such particles can still
interact with electromagnetic fields through nonminimal coupling, and
can be well described by the Dirac--Pauli equation. On the one hand,
without the minimal coupling, the Dirac--Pauli equation is simpler.
On the other hand, the interaction solely from nonminimal coupling
has not attracted enough attention, especially before
the discovery of the Aharonov--Casher (AC) effect [3-5]. Since the
AC effect is a consequence of the nonminimal coupling and has
been observed in experiment [5], one may become interested in
other consequences of the interaction. For instance, it may be of
interest to study bound states of neutral fermions in external
electromagnetic fields, especially when exact solutions are available.
It appears that this problem was not
considered in the literature as far as we know. The purpose of
this paper is to deal with this problem. It is organized as follows.

In the next section we consider the Dirac--Pauli equation of a neutral
fermion of spin $\frac 12$, with mass $m_{\rm n}$ and magnetic moment
$\mu_{\rm n}$, interacting with an external electromagnetic field
through nonminimal coupling. For spherically symmetric or central
electromagnetic fields, it can be shown that the total angular
momentum is a constant of motion. By separation of variables in
spherical coordinates, the stationary Dirac--Pauli
equation in a central
electric field, which involves four partial differential equations,
can be reduced to a
system of two coupled ordinary differential equations (ODE) for
two radial wave functions. Given a specific electric field, one
can in principle solve the system of ODE to obtain the radial
wave functions and determine the energy levels for bound states.
For a wide variety of electric fields, one can find bound-state
solutions of critical energy value $m_{\rm n}$ or $-m_{\rm n}$
in analytic forms.
It turns out that these critical energy levels are infinitely
degenerate. This is interesting because it reveals the possibility
of condensing infinitely many fermions, say, neutrons, into a
single energy level. Electric fields that  support a finite
number of critical bound states are also discussed.
In Sec. III we study a radially constant field, in this case the
system of ODE can be completely
solved, and we have scattering as well as bound-state solutions.
All bound-state solutions are given in closed forms.
Only the critical energy level has infinite degeneracy.
In Sec. IV we deal with radially linear
electric fields. The system of ODE is also completely solvable.
In this case we have only bound-state solutions, and many of the
energy levels are infinitely degenerate.
In Sec. V we discuss the simultaneous presence of central magnetic
and electric fields. In this case separation of variables is
still possible in spherical coordinates. But the reduced system of
ODE involves four coupled equations for four radial wave functions,
and is thus much more difficult to solve. Some other remarks and
discussins, say, the nonrelativistic limit,
are also included in this section.

\section{Neutral fermions in central electric fields}     

We work in (3+1)-dimensional space-time and use the natural units
where $\hbar=c=1$. Consider a neutral fermion of spin $\frac 12$
with mass $m_{\rm n}$ and magnetic monment $\mu_{\rm n}$,
moving in an external
electromagnetic field described by the field strength $F_{\mu\nu}$.
The fermion is described by a four-component spinorial wave function
$\Psi$ obeying the Dirac--Pauli equation [2, 6]
\bb
(i\gamma^\mu\partial_\mu-\textstyle{\frac 12}\mu_{\rm n}
\sigma^{\mu\nu}F_{\mu\nu}-m_{\rm n})\Psi=0,
\ee            
where $\gamma^\mu=(\gamma^0, {\bbox\gamma})$ are Dirac matrices
satisfying
\bb
\{\gamma^\mu,\gamma^\nu\}=2g^{\mu\nu}
\ee     
with $g^{\mu\nu}={\rm diag}(1,-1,-1,-1)$, and
\bb
\sigma^{\mu\nu}={i\over 2}[\gamma^\mu, \gamma^\nu].
\ee               
It can be shown that
\bb
\textstyle{\frac 12}\sigma^{\mu\nu}F_{\mu\nu}=
i{\bbox\alpha}\cdot{\bf E}-{\bbox\Sigma}\cdot{\bf B}
\ee             
where {\bf E} is the external electric field and {\bf B} the
magnetic one,
${\bbox\alpha}=\gamma^0{\bbox\gamma}$, and $\Sigma^k=\frac 12\epsilon
^{kij}\sigma^{ij}$ where $\epsilon^{kij}$ is totally antisymmetric
in its indices and $\epsilon^{123}=1$. If both {\bf E} and {\bf B} are
independent of the time $t$, one may set
\bb
\Psi(t,{\bf r})=e^{-i{\cal E}t}\psi({\bf r}),
\ee             
and obtain a stationary Dirac--Pauli equation for $\psi$:
$$
H\psi={\cal E}\psi,\eqno(6{\rm a})$$
where the Halmiltonian $H$ is given by
$$
H={\bbox\alpha}\cdot{\bf p}+i\mu_{\rm n}{\bbox\gamma}\cdot{\bf E}-
\mu_{\rm n}\gamma^0{\bbox\Sigma}\cdot{\bf B}+\gamma^0 m_{\rm n},
\eqno(6{\rm b})$$
where ${\bf p}=-i\nabla$ or $p^k=-i\partial_k$.

Now let us consider spherically symmetric or central fields
\addtocounter{equation}{1}
\bb
{\bf E}=E(r){\bf e}_r,\quad {\bf B}=B(r){\bf e}_r,
\ee             
where $r$ is one of the spherical coordinates $(r,\theta,\phi)$ and
${\bf e}_r$ is the unit vector in the radial direction. As usual we
define the the orbital angular momentum ${\bf L=r\times p}$. It is
not difficult to calculate the commutator $[{\bf L},H]$ and it turns
out that $[{\bf L},H]\ne 0$ even for free particles. For central
fields, however, it can be shown that the total angular momentum
${\bf J=L+S}$ where ${\bf S}=\frac 12{\bbox\Sigma}$ is a conserved
quantity, i.e., $[{\bf J},H]=0$. Thus one can have a common set of
eigenstates for $(H,{\bf J}^2,J_z)$. Because ${\bf S}^2=\frac 34$
is a constant operator, it is also a conserved quantity.
Unfortunately, ${\bf L}^2$ is not conserved and cannot have a common
set of eigenstates with $(H,{\bf J}^2,J_z)$. If $B(r)=0$, we have a
further conserved quantity $K=\gamma^0({\bbox\Sigma}\cdot{\bf L}+1)$
which commutes with both $H$ and {\bf J}. In this case one can
have a common set of eigenstates for $(H,{\bf J}^2,J_z,K,{\bf S}^2)$.

To solve the Dirac--Pauli equation, one should choose a specific
representation for the $\gamma$ matrices. Here we use the Dirac
representation [6]. In this representation we have ${\bbox\Sigma}=
{\rm diag}({\bbox\sigma},{\bbox\sigma})$ where ${\bbox\sigma}$ are
Pauli matrices, and ${\bf J}={\rm diag}({\bf j},{\bf j})$ where
${\bf j}={\bf L}+\frac 12{\bbox\sigma}$ is a $2\times 2$ matrix.
In this section we only consider a central electric field. The
simultaneous presence of a central magnetic field will be discussed
in Sec. V. We define $\psi=(\varphi,\chi)^\tau$, where $\tau$ denotes
transpose, and both $\varphi$ and $\chi$ are two-component spinors.
The stationary Dirac--Pauli equation (6) then takes the form
$$
{\bbox\sigma}\cdot({\bf p}-i\mu_{\rm n} E{\bf e}_r)\varphi
=({\cal E}+m_{\rm n})\chi,\eqno(8{\rm a})$$
$$
{\bbox\sigma}\cdot({\bf p}+i\mu_{\rm n} E{\bf e}_r)\chi
=({\cal E}-m_{\rm n})\varphi.\eqno(8{\rm b})$$
Here four partial differential equations are involved. We are going
to simplify these equations by separation of variables in spherical
coordinates. Let us define the two-component spinors
$$
f^+_{lm}(\theta, \phi)=\left(
\ba{c}\sqrt{\displaystyle {l+m+1\over 2l+1}}Y_{lm}(\theta, \phi)\\ 
\sqrt{\displaystyle {l-m\over 2l+1}}Y_{l,m+1}(\theta, \phi)
\ea
\right),\quad l=0,1,2,\ldots;m=-(l+1),-l,\ldots,l;
\eqno(9{\rm a})$$
$$
f^-_{lm}(\theta, \phi)=\left(
\ba{c}\sqrt{\displaystyle {l-m+1\over 2l+3}}Y_{l+1,m}(\theta, \phi)\\  
-\sqrt{\displaystyle {l+m+2\over 2l+3}}Y_{l+1,m+1}(\theta, \phi)
\ea
\right),\quad l=0,1,2,\ldots;m=-(l+1),-l,\ldots,l.
\eqno(9{\rm b})$$
Here $Y_{lm}(\theta, \phi)$ are spherical harmonics as defined in
Ref. [7]. Both of them are common eigenstates of
$({\bf j}^2,j_z,{\bf L}^2,{\bf S}^2)$ with eigenvalues
$$
((l+\textstyle{\frac 12})(l+\textstyle{\frac 32}),
m+\textstyle{\frac 12},l(l+1),\textstyle{\frac 34})$$
and
$$\textstyle{((l+\frac 12)(l+\frac 32),m+\frac 12,
(l+1)(l+2),\frac 34)},$$
respectively. It can be shown that
\addtocounter{equation}{2}
\bb
{\bbox\sigma}\cdot {\bf e}_r f^\pm_{lm}(\theta, \phi)
=f^\mp_{lm}(\theta, \phi),
\ee             
and
$$
{\bbox\sigma}\cdot {\bf L}f^+_{lm}(\theta, \phi)
=lf^+_{lm}(\theta, \phi),\eqno(11{\rm a})$$
$$
{\bbox\sigma}\cdot {\bf L}f^-_{lm}(\theta, \phi)
=-(l+2)f^-_{lm}(\theta, \phi).\eqno(11{\rm b})$$
The relation
\addtocounter{equation}{1}
\bb
{\bbox\sigma}\cdot {\bf p}=-i({\bbox\sigma}\cdot {\bf e}_r)\partial_r+
\frac ir({\bbox\sigma}\cdot {\bf e}_r)({\bbox\sigma}\cdot {\bf L})
\ee             
is also useful in the following. With these preparations we can
simplify Eq. (8) for two different kinds of solutions.

The first kind of solution to Eq. (8) is $\psi^+=(\varphi^+,\chi^+)^
\tau$, where
\bb
\varphi^+(r,\theta, \phi)
=u^+(r)f^+_{lm}(\theta, \phi),\quad
\chi^+(r,\theta, \phi)
=iv^+(r)f^-_{lm}(\theta, \phi).
\ee             
Note that $\psi^+$ is a  common eigenstate of
$({\bf J}^2,J_z,K,{\bf S}^2)$ with eigenvalues
$((l+\frac 12)(l+\frac 32),m+\frac 12,l+1,\frac 34)$, 
but it is not an eigenstate of ${\bf L}^2$. Using the relations
(10-12), it is not difficult to show that Eq. (8) now reduces to a
system of first-order ODE for the radial wave functions
$u^+(r)$ and $v^+(r)$:
$$
{du^+\over dr}+\mu_{\rm n}Eu^+-{l\over r}u^+=-({\cal E}+m_{\rm n})v^+,
\eqno(14{\rm a})$$
$$
{dv^+\over dr}-\mu_{\rm n}Ev^++{l+2\over r}v^+=({\cal E}
-m_{\rm n})u^+. \eqno(14{\rm b})$$
Because $\theta$ and $\phi$ are not defined at the origin, the
appropriate boundary conditions for $u^+$ and $v^+$ are
$$
|u^+(0)|<\infty\quad (l=0),\quad u^+(0)=0\quad (l\ne 0),
\eqno(15{\rm a})$$
$$
v^+(0)=0 \quad \forall l.\eqno(15{\rm b})$$
Of course they should also satisfy appropriate boundary conditions
at infinity. For bound-state solutions to be considered below, they
should fall off rapidly enough when $r\to \infty$ such that $\psi^+$
is square integrable. For scattering problem, they should be finite
at infinity. Given a specific form for $E(r)$, one can solve Eq. (14)
at least numerically. This is much simpler than dealing with Eq. (8).
For ${\cal E}\ne-m_{\rm n}$, one can express $v^+$ in terms
of $u^+$ by using Eq. (14a), and substitute it into Eq. (14b) to
obtain a second-order ODE solely for $u^+$:
\addtocounter{equation}{2}
\bb
{d^2u^+\over dr^2}+\frac 2r{du^+\over dr}+\left[{\cal E}^2-m_{\rm n}^2
+\mu_{\rm n} {dE\over dr}-\mu_{\rm n}^2E^2+2(l+1)\mu_{\rm n}\frac Er
-{l(l+1)\over r^2}\right]u^+=0.
\ee             
This is similar to the radial Schr\"odinger equation in a central
potential. It can be exactly solved for some specific form of $E(r)$.
This will be studied in the subsequent sections. When Eq. (16) is
solved, it is easy to obtain $v^+$. If ${\cal E}=-m_{\rm n}$,
one can directly solve Eq. (14) without difficulty.

The second kind of solution to Eq. (8) is $\psi^-=(\varphi^-,\chi^-)^
\tau$, where
\bb
\varphi^-(r,\theta, \phi)
=u^-(r)f^-_{lm}(\theta, \phi),\quad
\chi^-(r,\theta, \phi)
=iv^-(r)f^+_{lm}(\theta, \phi).
\ee             
Note that $\psi^-$ is also a  common eigenstate of
$({\bf J}^2,J_z,K,{\bf S}^2)$ with eigenvalues
$((l+\frac 12)(l+\frac 32),m+\frac 12,-(l+1),\frac 34)$, 
As before, Eq. (8) reduces to a
system of first-order ODE for the radial wave functions
$u^-(r)$ and $v^-(r)$:
$$
{du^-\over dr}+\mu_{\rm n}Eu^-+{l+2\over r}u^-=-({\cal E}
+m_{\rm n})v^-, \eqno(18{\rm a})$$
$$
{dv^-\over dr}-\mu_{\rm n}Ev^--{l\over r}v^-=({\cal E}-m_{\rm n})u^-.
\eqno(18{\rm b})$$
This is similar to Eq. (14).
If ${\cal E}\ne m_{\rm n}$, one can solve Eq. (18b) for $u^-$, and
substitute it into Eq. (18a) to obtain a
second-order ODE solely for $v^-$:
\addtocounter{equation}{1}
\bb
{d^2v^-\over dr^2}+\frac 2r{dv^-\over dr}+\left[{\cal E}^2-m_{\rm n}^2
-\mu_{\rm n} {dE\over dr}-\mu_{\rm n}^2E^2-2(l+1)\mu_{\rm n}\frac Er
-{l(l+1)\over r^2}\right]v^-=0.
\ee             
This is similar to Eq. (16). Note that the
appropriate boundary conditions for $u^-$ and $v^-$ at the origin are
$$
u^-(0)=0 \quad \forall l,\eqno(20{\rm a})$$
$$
|v^-(0)|<\infty\quad (l=0),\quad v^-(0)=0\quad (l\ne 0).
\eqno(20{\rm b})$$
Thus Eqs. (16) and (19) have the same boundary conditions at
the origin. Also note that they interchange if $E(r)\to -E(r)$.
If ${\cal E}=m_{\rm n}$, Eq. (19) is invalid, and one can solve
Eq. (18) directly.

Using the completeness relation of the spherical harmonics, it can
be shown that the two-component spinors $f^+_{lm}(\theta, \phi)$
and $f^-_{lm}(\theta, \phi)$ constitute a complete set on the sphere.
More specifically, we have
\addtocounter{equation}{1}
\bb
\sum_{l=0}^\infty\sum_{m=-(l+1)}^l[
f^+_{lm}(\theta, \phi)f^{+\dagger}_{lm}(\theta', \phi')+
f^-_{lm}(\theta, \phi)f^{-\dagger}_{lm}(\theta', \phi')]
=\delta(\cos\theta-\cos\theta')\delta(\phi-\phi').
\ee             
Therefore all possible forms of solutions to Eq. (8) are included in
Eqs. (13) and (17).

In the subsequent sections we are going to solve Eqs. (16) and (19)
for radially constant and radially linear electric fields. Before
dealing with these specific cases, we would like to give some special
bound-state solutions for more general forms of the electric field.
We assume that $E(r)$ behaves like $r^{-1+\delta_1}$ when $r\to 0$
and like $r^{-1+\delta_2}$ when $r\to\infty$ where $\delta_1$ and
$\delta_2$ are positive numbers, and is regular everywhere except
possibly at $r=0$. If $\mu_{\rm n}E(r)>0$ for $r>r_+$ where
$r_+$ is some finite radius, we have the following solution to
Eq. (14) with energy level ${\cal E}=m_{\rm n}$.
\bb
u^+_l(r)=A^+_l r^l\exp\left[-\int_0^r \mu_{\rm n} E(r')dr'\right],
\quad v^+_l(r)=0,
\ee             
where $A^+_l$ is a normalization constant. This obviously satisfies
the boundary conditions (15) at the origin. It is a bound-state
solution because it falls off rapidly enough to be square integrable.
It is remarkable that the energy eigenvalue does not depend on the
quantum numbers $l$ and $m$ (or $j=l+\frac 12$ and $m_j=m+\frac 12$).
Thus the degeneracy of this energy level is numerably infinite.
This is somewhat similar to the situation of a charged particle
in a magnetic field with infinite flux in two dimensions [8].
To our knowledge other similar  situations  were not encountered
previously in the realistic three-dimensional space.
This is interesting because it reveals the possibility of condensing
infinitely many fermions, say, neutrons, into a single energy level.
If $\mu_{\rm n}E(r)<0$ for $r>r_-$ where $r_-$ is some finite radius,
then we have the following solution to Eq. (18) with energy
level ${\cal E}=-m_{\rm n}$.
\bb
u^-_l(r)=0, \quad
v^-_l(r)=A^-_l r^l\exp\left[\int_0^r \mu_{\rm n} E(r')dr'\right],
\ee             
where $A^-_l$ is a normalization constant. This is also a bound-state
solution, and the energy level is infinitely degenerate. Here we have
a negative-energy solution, and will have more in the following
sections. The presence of negative-energy eigenvalues and
eigenstates is a quite general feature of relativistic quantum
mechanics. Though these solutions are nonphysical in the
one-particle theory, it is well known that they correspond to
antiparticles after second quantization. Therefore we will not
exclude these solutions in this paper.

If $E(r) \sim \kappa/r$ for large $r$ where $\kappa$ is a constant,
the situation is of special interest. In this case the nonvanishing
component of the critical solutions [$u_l^+(r)$ for $\mu_{\rm n}
\kappa>0$ or $v_l^-(r)$ for $\mu_{\rm n}\kappa<0$] does not fall off
exponentially at large $r$, but behaves like $r^{l-|\mu_{\rm n}
\kappa|}$. To be normalizable, one should have $l<|\mu_{\rm n}\kappa|
-\frac 32$. Therefore to have at least one critical bound state,
one should have $|\mu_{\rm n}\kappa|>\frac 32$. When $|\mu_{\rm n}
\kappa|-\frac 32$ is a natural
number, we have $|\mu_{\rm n}\kappa|-\frac 32$ critical boumd states
(degeneracy over $m$ has not been taken into account). When
$|\mu_{\rm n}\kappa|-\frac 32$ is not a natural number, the number of
critical bound states is $[|\mu_{\rm n}\kappa|-\frac 12]$, where the
square bracket denotes the integral part of a number.
This is similar to the situation of a charged particle
in a magnetic field with finite flux in two dimensions [8].

For a specific electric field, critical bound states with ${\cal E}
=m_{\rm n}$ and those with ${\cal E}=-m_{\rm n}$ do not appear
simultaneously.
This spectral asymmetry is also similar to that of a charged
particle in a magnetic field in two dimensions. Therefore vacuum
polarization similar to those in two dimensions for charged
particles [9-11] or neutral ones [12] may be expected for the
present system after second quantization.

To conclude this section we write down the normalization condition
for bound-state (or so called square integrable) solutions:
\bb
\int d{\bf r}\,\psi^{\pm\dagger}({\bf r})\psi^\pm({\bf r})=
\int_0^\infty [u^\pm(r)]^2r^2\,dr+
\int_0^\infty [v^\pm(r)]^2r^2\,dr=1.
\ee             
The normalization constants in Eqs. (22) and (23), and those in the
following sections are to be determined by this condition.

\section{Radially constant electric fields}   

In this section we consider radially constant electric fields
$E(r)=E_0$ where $E_0$ is a constant. As pointed out before, Eqs.
(16) and (19) interchange when $E(r)\to -E(r)$. So we need only
consider a positive $E_0$ or a negative $E_0$. For convenience we
assume that $\mu_{\rm n} E_0>0$. Now Eq. (16) takes the form
\bb
{d^2u^+\over dr^2}+\frac 2r{du^+\over dr}+\left[{\cal E}^2-m_{\rm n}^2
-\mu_{\rm n}^2E_0^2+{2(l+1)\mu_{\rm n} E_0\over r}-{l(l+1)\over r^2}
\right]u^+=0.
\ee             
This has the same form as the radial Schr\"odinger equation in an
attractive Coulomb field. The difference is that the ``Coulomb
field'' (the next to the last term in the square bracket) here depends
on the quantum number $l$. Thus the energy levels will depend on $l$
as well as a principal quantum number or a radial quantum number.
As Eq. (25) is familiar in quantum mechanics, we will give the
solutions only. Remember that Eq. (25) is invalid for ${\cal E}
=-m_{\rm n}$. We have the bound-state energy levels
$$
{\cal E}_{0l}=m_{\rm n},\quad n_r=0,
\eqno(26{\rm a})$$
$$
{\cal E}_{n_rl\pm}=\pm\left[m_{\rm n}^2+\mu_{\rm n}^2 E_0^2
{(n_r+l+1)^2
-(l+1)^2\over(n_r+l+1)^2}\right]^{\frac 12},\quad n_r=1,2,\ldots.
\eqno(26{\rm b})$$
Here $n_r$ is a radial quantum number. When $n_r=0$ we have a positive
critical energy level given in Eq. (26a). Though it is independent of
$l$, we keep the subscript $l$ to make a clear correspondence to the
corresponding wave functions below. For $n_r\ne 0$ we have positive-
and negative-energy levels, indicated by the subscript $\pm$ in Eq.
(26b). The corresponding radial wave functions are
$$
u^+_{n_rl\pm}(r)=A_{n_rl\pm}\rho^l e^{-\rho/2} L_{n_r}^{2l+1}(\rho),
\eqno(27{\rm a})$$
$$
v^+_{n_rl\pm}(r)=A_{n_rl\pm}{\mu_{\rm n} E_0\over (n_r+l+1)
({\cal E}_{n_rl\pm}+m_{\rm n})}\rho^{l+1} e^{-\rho/2}
L_{n_r-1}^{2l+3}(\rho)
\eqno(27{\rm b})$$
for $n_r\ne 0$, and
$$
u^+_{0l}(r)=A_{0l}\rho^l e^{-\rho/2}, \eqno(27{\rm c})$$
$$
v^+_{0l}(r)=0
\eqno(27{\rm d})$$
for $n_r=0$, where
\addtocounter{equation}{2}
\bb
\rho=\alpha_{n_rl}r,\quad
\alpha_{n_rl}={2(l+1)\mu_{\rm n} E_0\over (n_r+l+1)},
\ee             
and $L_{n_r}^{2l+1}(\rho)$, etc., are Laguerre polynomials defined in
Ref. [13], which are different from those used in Ref. [7]. Note that
the superscript + indicates the first kind of solutions (13), while
the subscript $\pm$ indicates the sign of the energy levels. It is
seen from Eq. (27) that negative- and positive-energy solutions have
the same functional form, but the coefficients are different.
The normalization constants are
$$
A_{n_rl\pm}={(\mu_{\rm n} E_0)^{\frac 32}\over (n_r+l+1)^2}
\left[{2(l+1)^3 n_r!\over (n_r+2l+1)!}\right]^{\frac 12}
\left({{\cal E}_{n_rl\pm}+m_{\rm n}\over {\cal E}_{n_rl\pm}}\right)
^{\frac 12}
\eqno(29{\rm a})$$
for $n_r\ne 0$ and
$$
A_{0l}={(2\mu_{\rm n} E_0)^{\frac 32}\over\sqrt{(2l+2)!}}
\eqno(29{\rm b})$$
for $n_r=0$. The degeneracy of the energy level ${\cal E}_{n_rl+}$
or ${\cal E}_{n_rl-}$ is $2l+2$. As the energy level ${\cal E}_{0l}
=m_{\rm n}$ is actually independent of $l$, its degeneracy is
numerably infinite. Indeed, the solution (28) is a specific
realization of the solution (22) discussed before.

When ${\cal E}=-m_{\rm n}$, Eq. (25) is invalid. In this case one
should deal with Eq. (14) directly. It is easy to show that this
energy value corresponds to a trivial solution. Thus all first-kind
solutions are included in Eq. (27), and the corresponding
energy levels are given by Eq. (26). Note that all energy levels have
absolute values less than $\sqrt{m_{\rm n}^2+\mu_{\rm n}^2E_0^2}$.
When $|{\cal E}|$ exceeds this value, we have scattering solutions to
Eq. (25). This will not be discussed here.

Now we turn to Eq. (19), which in the present case becomes
\addtocounter{equation}{1}
\bb
{d^2v^-\over dr^2}+\frac 2r{dv^-\over dr}+\left[{\cal E}^2-m_{\rm n}^2
-\mu_{\rm n}^2E_0^2-{2(l+1)\mu_{\rm n} E_0\over r}-{l(l+1)\over r^2}
\right]v^-=0.
\ee             
Since $\mu_{\rm n} E_0>0$, this is equivalent to
the radial Schr\"odinger equation in a repulsive Coulomb field.
In this case only scattering solutions are available.
These scattering solutions have energy
${\cal E}>\sqrt{m_{\rm n}^2+\mu_{\rm n}^2E_0^2}$ or
${\cal E}<-\sqrt{m_{\rm n}^2+\mu_{\rm n}^2E_0^2}$.
If ${\cal E}=m_{\rm n}$, Eq. (30) is invalid. Then we may
deal with Eq. (18) directly. It turns out that this energy
value corresponds to a trivial solution. We thus conclude that 
there is no bound state of the second kind in the present case.

To finish this section we estimate the ``Bohr radius'' of the neutron
in this radially constant field. It is roughly equal to
$\alpha_{n_r l}^{-1}$. For the critical-energy state, $n_r=0$, and
$\alpha_{0l}^{-1}=(2\mu_{\rm n}E_0)^{-1}$. In the MKS system it reads
\bb
\alpha_{0l}^{-1}={\hbar c^2\over 2\mu_{\rm n}E_0}.
\ee             
We take $|E_0|=5.15\times 10^{11}$ V/m, the electric field strength
at the Bohr radius of the hydrogen atom. For neutrons, we have
$\alpha_{0l}^{-1}=9.5\times 10^{-4}$ m. This is a macroscopic length
scale but rather small. However, it might be not easy to realize a
radially constant central electric field with the above magnitude in
the laboratory. We do not know whether there exists some such field
somewhere in the universe.

\section{Radially linear electric fields}  

In this section we turn to another exactly solvable field, the
radially linear electric field $E(r)=\beta r$ where $\beta$ is a
constant. The electric charge density that produces this field
is $\rho_{\rm c}=3\beta/4\pi$ in the Gaussian units, which is a
constant. To realize the above central field, however, the electric
charge density should become zero outside some large sphere where the
particle under consideration cannot reach practically. Otherwise the
electric field would be zero everywhere. In the region of interest
(inside the large sphere) the field is then radially linear. For
reasons given earlier, we need only consider one sign of $\beta$.
For convenience we assume that $\beta\mu_{\rm n}>0$. Eq. (16)
then takes the form
\bb
{d^2u^+\over dr^2}+\frac 2r{du^+\over dr}+\left[{\cal E}^2-m_{\rm n}^2
+(2l+3)\beta\mu_{\rm n}-\beta^2\mu_{\rm n}^2r^2 -{l(l+1)\over r^2}
\right]u^+=0.
\ee             
This is not valid for ${\cal E}=-m_{\rm n}$. In the latter case one
can solve Eq. (14) directly and obtain a trivial solution. Thus all
nontrivial solutions of the first kind are those arise from Eq. (32).
The equation (32) has the same form as the radial Schr\"odinger
equation for an isotropic harmonic oscillator. The difference is that
the ``energy'' here depends on the quantum number $l$. Thus the
dependence of the energy levels on the quantum numbers will be
different from that of the isotropic harmonic oscillator.
Since Eq. (32)
is also familiar in quantum mechanics, we will give the solutions
only. There are only bound-state solutions. The energy levels are
$$
{\cal E}_0^+=m_{\rm n}, \quad n_r=0
\eqno(33{\rm a})$$
$$
{\cal E}_{n_r\pm}^+=\pm \sqrt{m_{\rm n}^2+4n_r\beta\mu_{\rm n}}, \quad
n_r=1,2,\ldots,
\eqno(33{\rm b})$$
where $n_r$ is a radial quantum number. Note that the superscript
+ for ${\cal E}$ indicates the first kind of solutions, while the
subscript $\pm$ indicates the sign of the energy. As before, we have
negative- as well as positive-energy levels.
The corresponding radial wave functions are
\addtocounter{equation}{1}
\bb
u^+_{0l}(r)=A^+_{0l}\rho^l e^{-\rho^2/2}, \quad
v^+_{0l}(r)=0
\ee            
for $n_r=0$, and
$$
u^+_{n_rl\pm}(r)=A^+_{n_rl\pm}\rho^l e^{-\rho^2/2}L_{n_r}^{l+1/2}
(\rho^2),
\eqno(35{\rm a})$$
$$
v^+_{n_rl\pm}(r)=A^+_{n_rl\pm}{2\sqrt{\beta\mu_{\rm n}}\over 
{\cal E}^+_{n_r\pm}+m_{\rm n}}\rho^{l+1} e^{-\rho^2/2}
L_{n_r-1}^{l+3/2}(\rho^2)
\eqno(35{\rm b})$$
for $n_r\ne 0$, where
\addtocounter{equation}{1}
\bb
\rho=\sqrt{\beta\mu_{\rm n}}r
\ee             
and $L_{n_r}^{l+1/2}(\rho^2)$, etc., are Laguerre polynomials as
employed in Sec. III but the argument here is $\rho^2$.
The normalization constants are determined by Eq. (24) and are given
by
$$
A^+_{0l}={\sqrt2(\beta\mu_{\rm n})^{\frac 34}\over\sqrt
{\Gamma(l+3/2)}},\quad n_r=0 \eqno(37{\rm a})$$
$$
A^+_{n_rl\pm}=(\beta\mu_{\rm n})^{\frac 34}
\left[{n_r!\over \Gamma(n_r+l+3/2)}\right]^{\frac 12}
\left({{\cal E}^+_{n_r\pm}+m_{\rm n}\over {\cal E}^+_{n_r\pm}}
\right)^{\frac 12},
\quad n_r=1,2,\ldots.
\eqno(37{\rm b})$$
It is remarkable that all the above energy levels are
independent of the quantum number $l$, and thus all of them are
infinitely degenerate.
The critical-energy solution (34) is another realization of the 
solution (22).

Now we consider the second kind of solutions (17). It is easy to
show that Eq. (18) gives a trivial solution when
${\cal E}=m_{\rm n}$. Thus all nontrivial solutions arise from Eq.
(19) which is valid for ${\cal E}\ne m_{\rm n}$ and in the
present case becomes
\addtocounter{equation}{1}
\bb
{d^2v^-\over dr^2}+\frac 2r{dv^-\over dr}+\left[{\cal E}^2-m_{\rm n}^2
-(2l+3)\beta\mu_{\rm n}-\beta^2\mu_{\rm n}^2r^2-{l(l+1)\over r^2}
\right]v^-=0.
\ee             
This is very similar to Eq. (32). The only difference lies in the
sign of the third term in the square bracket. This difference,
however, will render the energy levels quite different from those
obtained above. As before, we only give the results here.
The energy levels are
\bb
{\cal E}^-_{N\pm}=\pm \sqrt{m_{\rm n}^2+(4N+6)\beta\mu_{\rm n}}, \quad
N=n_r+l=0,1,2,\ldots,
\ee             
where $n_r=0,1,2,\ldots$ is a radial quantum number and $N$ is a
principal quantum number.
The superscript
$-$ of ${\cal E}$ indicates the second kind of solutions, while the
subscript $\pm$ indicates the sign of the energy.
The spectrum obtained here has no overlap with that in Eq. (33).
The corresponding radial wave functions are
$$
u^-_{n_rl\pm}(r)=-A^-_{n_rl\pm}{2\sqrt{\beta\mu_{\rm n}}\over 
{\cal E}^-_{N\pm}-m_{\rm n}}\rho^{l+1} e^{-\rho^2/2}
L_{n_r}^{l+3/2}(\rho^2),
\eqno(40{\rm a})$$
$$
v^-_{n_rl\pm}(r)=A^-_{n_rl\pm}\rho^l e^{-\rho^2/2}L_{n_r}^{l+1/2}
(\rho^2),
\eqno(40{\rm b})$$
where $\rho$ is given by Eq. (36), and $n_r=0,1,2,\ldots$
is the radial quantum number.
The normalization constants are given by
\addtocounter{equation}{1}
\bb
A^-_{n_rl\pm}=(\beta\mu_{\rm n})^{\frac 34}
\left[{n_r!\over \Gamma(n_r+l+3/2)}\right]^{\frac 12}
\left({{\cal E}^-_{N\pm}-m_{\rm n}\over {\cal E}^-_{N\pm}}\right)
^{\frac 12},
\quad n_r=0,1,2,\ldots.
\ee             
The energy levels ${\cal E}^-_{N+}$ and ${\cal E}^-_{N-}$ depend
only on the principal quantum number $N$. Given $N$, $l$ may vary
from 0 to $N$. For a given $l$, there are $2l+2$ different solutions.
Therefore the degeneracy of the level
${\cal E}^-_{N+}$ or ${\cal E}^-_{N-}$ is
\bb
d_N=\sum_{l=0}^N (2l+2)=(N+1)(N+2).
\ee             

In conclusion, in the radially linear electric field, we have two
sets of bound-state energy levels. The first set is given in Eq. (33),
corresponding to the first kind of solutions. 
The second set is given in Eq. (39),
corresponding to the second kind of solutions. There is no scattering
solution here. In contrast, the radially constant electric field
studied in Sec. III admits both scattering and bound-state solutions,
though there exists no bound state of the second kind. Finally we
estimate the ``Bohr radius'' of the neutron in the present case.
This is roughly equal to $(\beta\mu_{\rm n})^{-\frac 12}$, or
$(3/4\pi\rho_{\rm c}\mu_{\rm n})^{\frac 12}$ where $\rho_{\rm c}$ is
the electric  charge density producing the field. In the MKS system
this reads
$$
\left(3\hbar\over 4\pi\mu_0\rho_{\rm c}\mu_{\rm n}\right)
^{\frac 12},$$
where $\mu_0$ is the permeability of the vacuum. We take
$\rho_{\rm c}=e/a_0^3$ where $e$ is the electron charge and
$a_0$ is the Bohr radius of the hydrogen atom. For neutrons the
above ``Bohr radius'' has the value $4.4\times 10^{-8}$ m. This is
rather small. However, it may be difficult to achieve the above
electric charge density.

\section{Summary and discussions}  

In the preceding sections we have studied the Dirac--Pauli equation of a
neutral fermion with nonminimal coupling to a central electric field.
By separation of variables in spherical coordinates, the stationary
Dirac--Pauli equation which involves four partial differential equations
can be reduced to a system of ODE which involves two
coupled first-order ODE
for two radial wave functions. There are two different kinds of
solutions, and thus two independent systems of ODE.
Bound states of critical energy values can be obtained analytically
for a quite general class of electric fields, where the degeneracy
of the critical energy level turns out to be numerably infinite.
This reveals the possibility
of condensing infinitely many fermions into a
single energy level.
We also discussed a special form of the electric field that  supports
a finite number of critical bound states.
Two specific electric fields, one radially
constant and the other radially linear, are studied in detail and
all the bound-state solutions are obtained in closed forms.
In the first case bound states exist only for the first kind of
solutions, while scattering states exist for both kinds.
Scattering states
are not discussed in detail. In the second case, we have two sets
of discrete energy levels corresponding to the two kinds of
solutions. There is no scattering state.
It turns out that the energy levels in the first set are all 
infinitely degenerate. In both fields we have negative as well as
positive energy levels.
Critical energy levels are also admitted in both cases, which
may be positive or negative depending on the signs of $\mu_{\rm n}$
and the electric fields. Note that the two critical energy levels
are not admitted at the same time, however. This spectral
asymmetry may likely lead to vacuum polarization after second
quantization.

In Sec. II we have shown that the total angular momentum {\bf J} is
a conserved quantity in the simultaneous presence of a central
magnetic field and a central electric field. But we have not discussed
the solutions of the Dirac--Pauli equation in this case. In the Dirac
representation, the stationary Dirac--Pauli equation (6) takes the form
$$
{\bbox\sigma}\cdot({\bf p}-i\mu_{\rm n} E{\bf e}_r)\varphi
=({\cal E}+m_{\rm n}-\mu_{\rm n} B{\bbox\sigma}\cdot{\bf e}_r)\chi,
\eqno(43{\rm a})$$
$$
{\bbox\sigma}\cdot({\bf p}+i\mu_{\rm n} E{\bf e}_r)\chi
=({\cal E}-m_{\rm n}+\mu_{\rm n} B{\bbox\sigma}\cdot{\bf e}_r)\varphi.
\eqno(43{\rm b})$$
These equations are similar to Eq. (8) but more complicated. They
are still separable in spherical coordinates. We set
$$
\varphi(r,\theta, \phi)
=u^+(r)f^+_{lm}(\theta, \phi)+u^-(r)f^-_{lm}(\theta, \phi),
\eqno(44{\rm a})$$
$$
\chi(r,\theta, \phi)
=iv^+(r)f^-_{lm}(\theta, \phi)+iv^-(r)f^+_{lm}(\theta, \phi).
\eqno(44{\rm b})$$
Substituting these ansatz into Eq. (43) and using the relations
(10-12) we obtain the following system of ODE for the four radial
wave functions.
$$
{du^+\over dr}+\mu_{\rm n}Eu^+-{l\over r}u^+=-({\cal E}+m_{\rm n})v^+
+\mu_{\rm n}Bv^-,
\eqno(45{\rm a})$$
$$
{dv^+\over dr}-\mu_{\rm n}Ev^++{l+2\over r}v^+=({\cal E}-m_{\rm n})u^+
+\mu_{\rm n}Bu^-,
\eqno(45{\rm b})$$
$$
{du^-\over dr}+\mu_{\rm n}Eu^-+{l+2\over r}u^-=-({\cal E}+
m_{\rm n})v^-+\mu_{\rm n}Bv^+,
\eqno(45{\rm c})$$
$$
{dv^-\over dr}-\mu_{\rm n}Ev^--{l\over r}v^-=({\cal E}-m_{\rm n})u^-
+\mu_{\rm n}Bu^+.
\eqno(45{\rm d})$$
\addtocounter{equation}{3}
If $B(r)=0$, one may set $u^-=v^-=0$ which reduces Eq. (45) to Eq.
(14) for the first kind of solutions,
or set $u^+=v^+=0$ which reduces Eq. (45) to Eq.
(18) for the second kind of solutions. This is what we have done
before for a pure electric field. When a magnetic field is present
at the same time, this is not allowed, however.
The essential reason is that $K$ is no longer a conserved quantity
in this case. All the four ODE
in Eq. (45) are coupled to each other. It seems difficult to solve
them even for a pure magnetic field. We will not go into further
details of these equations here.

Let us briefly discuss the nonrelativistic limit  of the
Dirac--Pauli equation. Consider the stationary case  with a
pure electric field. We can solve Eq. (8a) for $\chi$, and substitute
it into Eq. (8b) to obtain an equation for $\varphi$:
\bb
[{\bbox\sigma}\cdot({\bf p}+i\mu_{\rm n} {\bf E})]
[{\bbox\sigma}\cdot({\bf p}-i\mu_{\rm n} {\bf E})]\varphi
=({\cal E}^2-m_{\rm n}^2)\varphi.
\ee     
This holds for any value of ${\cal E}$ except ${\cal E}=-m_{\rm n}$,
and is valid for noncentral electric field as well. To discuss the
nonrelativistic limit we consider only positive ${\cal E}$ and set
$$
{\cal E}=m_{\rm n}+{\cal E}'.
$$
When ${\cal E}'\ll m_{\rm n}$ we get the nonrelativistic limit of
Eq. (46):
\bb
[{\bbox\sigma}\cdot({\bf p}+i\mu_{\rm n} {\bf E})]
[{\bbox\sigma}\cdot({\bf p}-i\mu_{\rm n} {\bf E})]\varphi
=2m_{\rm n}{\cal E}'\varphi.
\ee     
This has essentially the same form as Eq. (46), and thus the same
solutions. However, it should be remarked that even when
$|\mu_{\rm n} {\bf E}|\ll m_{\rm n}$, Eq. (47) is not valid for those
${\cal E}'$ comparable with $m_{\rm n}$. For example, in the radially
constant field with $|\mu_{\rm n} E_0|\ll m_{\rm n}$, Eq. (47) is
good for all bound states, but not for scattering ones with large
${\cal E}$, say, ${\cal E}=2 m_{\rm n}$. On the other hand, even
if $|{\bf E}|$ is unbounded, Eq. (47) is still valid for small
${\cal E}'$. For example, in the radially linear field, Eq. (47)
may be good for lower levels if $|\beta\mu_{\rm n}|\ll m_{\rm n}$.
Since Eq. (47) is not simpler, it is more convenient to deal with
Eq. (46) directly. The nonrelativistic limit with both magnetic and
electric fields can be similarly discussed, though the situation is
more complicated. We will not give further details here.

We have pointed out in Sec. III that the radially constant electric
field admit scattering solutions of both kinds. Though Eqs. (25) and
(30) can be solved to give partial wave solutions, the scattering
problem is difficult to handle in this case since these equations
involve long-range ``Coulomb potentials''. An easier situation for
the scattering problem may be the field $E(r)\propto r^{-1}$.
This will be studied subsequently.

In this paper we have dealt with (3+1)-dimensional problems. The
Dirac--Pauli equation (1) has a much simpler form in a
(2+1)-dimensional
space-time. Indeed, the situation for the AC effect is equivalent
to a (2+1)-dimensional problem because of the specific field
configuration. Recently, we have calculated the probability of
neutral particle-antiparticle pair creation in the vacuum by
external electromagnetic fields in 2+1 dimensions, based on the
nonminimal coupling [14]. Both scattering and bound-state problems in
external fields are easier in 2+1 dimensions. These and other
consequences of the nonminimal coupling will also be
studied subsequently.

\section*{Acknowledgments}

The author is grateful to Professor Guang-jiong Ni for encouragement.
This work was supported by the
National Natural Science Foundation of China.


\newpage
\begin{thebibliography}{99}

\bibitem{1}W. Pauli, Rev. Mod. Phys. {\bf 13}, 203 (1941).
\bibitem{2}V. G. Bagrov and D. M. Gitman, {\it Exact Solutions of
Relativistic Wave Equations} (Kluwer, Dordrecht, 1990),
Chapter 7 and references therein.

\bibitem{3}Y. Aharonov and A. Casher, Phys. Rev. Lett. {\bf 53},
 319 (1984); A. S. Goldhaber, Phys. Rev. Lett. {\bf 62}, 482 (1989).

\bibitem{4}T. H. Boyer, Phys. Rev. A {\bf 36}, 5083 (1987);
Y. Aharonov, P. Pearle, and L. Vaidman, Phys. Rev. A {\bf 37},
4052 (1988).

\bibitem{5}C. Cimmino, G. I. Opat, A. G. Klein, H. Kaiser, S. A.
Werner, M. Arif, and R. Clothier, Phys. Rev. Lett. {\bf 63},
 380 (1989).

\bibitem{6}C. Itzykson and J.-B. Zuber, {\it Quantum Field Theory}
(McGraw-Hill, New York, 1980).

\bibitem{7}L. I. Schiff, {\it Quantum Mechanics}
(McGraw-Hill, New York, 1968).
\bibitem{8}Y. Aharonov and A. Casher, Phys. Rev. A {\bf 19},
 2461 (1979); R. Jackiw, Phys. Rev. D {\bf 29}, 2375 (1984).
\bibitem{9}A. P. Polychronakos, Nucl. Phys. B {\bf 278},
 207 (1986).
\bibitem{10}A. J. Niemi and G. W. Semenoff, Phys. Rev. Lett. {\bf 51},
 2077 (1983); Phys. Rep. {\bf 135}, 99 (1986).  
\bibitem{11}A. N. Redlich, Phys. Rev. Lett. {\bf 52},
 18 (1984); Phys. Rev. D {\bf 29}, 2366 (1984).
\bibitem{12}Q.-G. Lin, Zhongshan University preprint.

\bibitem{13}I. S. Gradshteyn and I. M. Ryzhik, {\it Tables of
Integrals, Series, and Products} (Academic, New York, 1980).
\bibitem{14}Q.-G. Lin, J. Phys. G {\bf 25}, 1793 (1999).
\end{thebibliography}
\end{document}